\documentclass{article}

\usepackage{color}
\usepackage{graphicx}
\usepackage{seqsplit}

\font\scap=cmcsc10

\font\tenmsb=msbm10

\newcount\eqnumber
\def\neweq{{\rm{(\the\eqnumber)}}\global\advance\eqnumber by 1}
\def\eqdef#1{\eqno\xdef#1{\the\eqnumber}\neweq}
\def\newaeq{{\rm{(\the\eqnumber a)}}\global\advance\eqnumber by 1}
\def\eqdaf#1{\eqno\xdef#1{\the\eqnumber}\newaeq}
\def\eqdisp#1{\xdef#1{\the\eqnumber}\neweq}
\def\eqdasp#1{\xdef#1{\the\eqnumber}\newaeq}

\newcount\refnumber
\def\newref{{\the\refnumber}\global\advance\refnumber by 1}
\def\refdef#1{{\xdef#1{\the\refnumber}}\newref}

\newcount\fignumber
\def\newfig{{\the\fignumber}\global\advance\fignumber by 1}
\def\figdef#1{{\xdef#1{\the\fignumber}}\newfig}

\def\smallskip{\vskip 3pt}
\def\medskip{\vskip 6pt}
\def\bigskip{\vskip 12pt}

\def\bol{\scalebox{0.7}{$\bullet$}}

\begin{document}

\centerline{\bf A fast algorithmic way to calculate the degree growth of birational mappings}
\bigskip
\medskip {\scap B. Grammaticos} and {\scap A. Ramani}
\quad{\sl Universit\'e Paris-Saclay and Universit\'e de Paris-Cit\'e, CNRS/IN2P3, IJCLab, 91405 Orsay, France}
\medskip{\scap A.S. Carstea}
\quad{\sl Department of Theoretical Physics, NIPNE, Magurele 077125, Romania}
\medskip{\scap R. Willox} \quad
{\sl Graduate School of Mathematical Sciences, the University of Tokyo, 3-8-1 Komaba, Meguro-ku, 153-8914 Tokyo, Japan }

\bigskip
{\sl Abstract}
\smallskip
We present an algorithmic method for the calculation of the degrees of the iterates of birational mappings, based on Halburd's method for obtaining the degrees from the singularity structure of the mapping. The method uses only integer arithmetic with 
additions and, in some cases, multiplications by small integers. It is therefore extremely fast. Several examples of integrable and non-integrable mappings are presented. In the latter case the dynamical degree we obtain from our method is always in agreement with that calculated by previously known methods. 
\bigskip

\bigskip
1. {\scap Introduction}
\medskip
Integrability is a rare phenomenon. This fact, linked to the nice properties it is generally associated with, led at the beginning of the 20th century to a profusion of studies [\refdef\ince] dealing almost exclusively with ordinary differential equations. Subsequently however, the interest in integrability waned considerably and its revival only came with the discovery of the soliton and of integrable partial differential equations in the 1960s [\refdef\martin]. The main bulk of the ensuing studies again concerned differential systems, but a notable exception were the works of Hirota [\refdef\hirota] who, singlehandedly, derived discrete analogues of all the classical integrable evolution equations. His studies remained mainly ignored till the 1990s, when the interest of the integrable community shifted towards discrete systems, giving rise to an explosion of results in the domain.

Put in very vague terms, integrability is associated with the existence of a `sufficient number' of conserved quantities. (The authors are of course aware that for the Painlev\'e equations, to which they have devoted several studies [\refdef\cimpa], the relation of integrability to conserved quantities is rather tenuous). While the existence of conserved quantities does (again, with the appropriate caveats) guarantee integrability, it is not of great use when one is facing the question whether a given system is integrable or not. What is really needed is an easily implementable integrability criterion. 

By analogy to the continuous, differential, systems, where singularities play a capital role in integrability [\refdef\physrep], it was considered that the structure of singularities of a discrete system could provide useful information on its integrable or non-integrable character. All the more so as it was observed that several systems, all of them integrable through spectral methods, had a very special singularity structure: a singularity that appeared at some iteration step, due to a special choice of initial conditions, disappeared again after a few more iterations. This property was dubbed singularity confinement [\refdef\sincon] and has been extensively used as a discrete integrability criterion ever since. For example, the derivation of the discrete analogues of the Painlev\'e equations [\refdef\dps] largely proceeded based on it. However, it turned out that the singularity confinement property is not a sufficient condition [\refdef\hiv] for integrability. In fact, infinitely many non-integrable discrete systems exist, having confined singularities. What was therefore needed was a more refined way of analysing the singularity confinement constraints. This was done in the approach we dubbed full-deautonomisation  [\refdef\redem], which allows one to distinguish between integrable and non-integrable discrete systems with confined singularities. A rigorous, algebro-geometric justification of the full-deautonomisation method was given in  [\refdef\stokes].

Another approach to the detection of discrete integrability is based on the formal identification of discrete systems and delay equations. The studies of Ablowitz and collaborators  [\refdef\ablow] have shown that at the continuum limit, when a delay equations goes over to a differential one, the singularities of the latter that are at finite distance can be associated to the behaviour of the solutions of the discrete system at infinity. What ensued was an approach aiming at studying this behaviour using the arsenal of the Nevanlinna theory  [\refdef\nevan]. We will not discuss further the success of the Nevanlinna approach for the characterisation of discrete integrability. The point we shall retain from these studies is that integrability is incompatible with too fast an asymptotic growth of the solutions of the discrete system.

Veselov  [\refdef\veselov] put this in a more general setting by stating that `[discrete] integrability has an essential correlation with the weak growth of certain characteristics', based on a statement by Arnold [\refdef\arnold] introducing the notion of complexity for mappings on the plane. The latter is defined as the number of intersection points of a fixed curve with the images of a second curve under the $n$-{th} iteration of the mapping. In fact, when it comes to mappings on the plane, Diller and Favre  [\refdef\diller] have presented a classification of the possible degree growths (something equivalent to Arnold's complexity) of the iterates of (autonomous) second-order birational autonomous mappings. Three integrable cases can be distinguished:

-- The degree growth is bounded. In this case the mapping is either periodic or can be transformed into a projective mapping after some birational transformations.

-- The degree grows linearly with $n$. In this case the mapping is linearisable as it preserves a rational fibration. Moreover, such a mapping has at least one non-confined singularity.

-- The degree grows quadratically with $n$. In this case the mapping is considered to be integrable as it always preserves an elliptic fibration and, moreover, it always enjoys the singularity confinement property.

Non-integrable mappings on the other hand have exponential degree growth, have no typical preserved structures but, as mentioned above, may or may not possess the singularity confinement property. 

A simple way to encode these different behaviours is through  the dynamical degree. If $d_n$ represents the degree of the $n$th iterate of a rational mapping, then the dynamical degree is defined as
$$\lambda=\lim_{n\to\infty}d_n^{1/n}.\eqdef\zena$$
Integrable mappings have a dynamical degree equal to 1, while a dynamical degree larger than 1 signals non-integrability. 

In a nutshell, if one wishes to make predictions concerning the integrable character of a discrete system it suffices to examine its complexity properties. For a rational mapping this is rather straightforward: introduce appropriate initial conditions, iterate them and compute the degree of a suitably chosen variable in the numerator and denominator of the rational expression obtained.  This paper is devoted to the calculation of the successive degrees obtained when iterating a birational mapping. A fast, algorithmic, approach is presented, based on the theory proposed by R. Halburd in [\refdef\rod]. In what follows we start with a short refresher on how the degrees are usually calculated and then we proceed to the presentation of our method.
\bigskip
2. {\scap A refresher}
\medskip
Suppose one is given a one-dimensional birational mapping and wishes to make a prediction concerning its integrability. The simplest way to do this is by computing its dynamical degree. At this point it should be stressed that there is no theorem like the one of Diller and Favre concerning mappings of order higher than two, so concluding that the mapping is not integrable once the dynamical degree turns out to be larger than 1 may sound somewhat arbitrary. Still, we believe this assumption to be a reasonable one. 

By far the simplest way to compute the dynamical degree of a mapping is by using the Diophantine method introduced by Halburd  [\refdef\dioph].
Here is how the method works. One starts from initial conditions that are simple rational numbers and then computes the successive fractions obtained by iterating the mapping. The calculations are simple, all the more so since one does not have to manipulate any symbolic variables. From the successive iterates of the mapping, obtained as fractions of the form $N_n/D_n$, one computes the successive approximations of the dynamical degree as $\log N_n/\log N_{n-1}$ (or equivalently $\log D_n/\log D_{n-1}$). If the mapping is integrable the dynamical degree converges (in general, very slowly) to 1. One typically needs hundreds of  additional iterations in order to gain an order of magnitude in accuracy, but the calculations in this case are rather fast (since the size of the fractions involved grows slowly). In the case of non-integrable mappings the convergence is much faster but, since the growth of the size of the fractions involved is exponential, the calculations can easily become prohibitive. Let us illustrate these points with two tangible examples.

We start with a well-known integrable mapping
$$x_{n+1}x_{n-1}=1+{1\over x_n}.\eqdef\zdyo$$
We choose some integer values for $x_0$ and $x_1$ and compute the dynamical degree using the Diophantine prescription. After 500 iterations we find $\lambda\approx1.004$ which becomes 1.002 after 1000 iterations, the calculation being 30 times longer this time. However it is clear that the value of the dynamical degree converges to 1, in agreement with the integrable character of the mapping. 

Next we examine the non-integrable mapping  [\refdef\hone]
$$x_{n+1}x_{n-1}=x_n+{1\over x_n}.\eqdef\ztri$$
In [\redem] we obtained the exact expression of its dynamical degree $(1+\sqrt{17})/4+\sqrt{(1+\sqrt{17})/8}$, with approximate value 2.081019. Using the Diophantine method we find that after 10 iterations only the first decimal after 2, i.e. 2.0,  has converged. Pushing the calculation to 20 iterations gives a value where four decimals have converged, i.e. 2.0810, but the time necessary for this calculation is a few thousand times longer. Of course, with four converged decimals one does not really need to push the calculation further. 

Similar behaviour is observed in every case, be it of order two or higher, that we have studied. The Diophantine method is a reliable tool for the obtention of the dynamical degree. However it does not provide access to the degrees themselves and can lead to prohibitively long calculations in the case of non-integrable mappings.
Finding the exact degree growth is in fact quite simple. We shall illustrate this using the two mappings introduced above, while also pointing out the relation of the growth to the singularity structure of the mapping. 

We start from (\zdyo) and take initial conditions $x_0=1$ and $x_1=z$. We then iterate the mapping and find the following sequence of iterates (where $P_n(z)$ represents a polynomial of degree $n$ in $z$)
$$x_2={z+1\over z}$$
$$x_3={2z+1\over z(z+1)}$$
$$x_4={zP_2(z)\over(z+1)(2z+1)}$$
$$x_5={(z+1)P_3(z)\over (2z+1)P_2(z)}$$
$$x_6={(z+1) (2z+1)P_4(z)\over z(z+1)P_2(z)P_3(z)}$$
$$x_7={(z+1)^2 (2z+1)P_2(z)P_5(z)\over (z+1)^2 (2z+1)P_3(z)P_4(z)}$$
$$x_8={z(z+1)^3 (2z+1)^2P_2(z)P_3(z)P_6(z)\over (z+1)^3 (2z+1)^2P_2(z)P_4(z)P_5(z)},$$
where we have yet to implement all possible simplifications between the numerators and denominators in the above rational functions.
The degrees computed as the maximum of the degrees in $z$ of these numerators and denominators, before simplification, are 0, 1, 1, 2, 3, 4, 7, 10, 17. Once the simplifications are carried out however, the degrees become 0, 1, 1, 2, 3, 4, 6, 7, 10.  In order to understand how these simplifications come about we must discuss the singularities and indeterminacies of the mapping. Suppose that $z$ takes the value $-1$.  Then $x_2$ vanishes and the  successive values of $x_n$ are $-1,0,\infty,\infty,0,-1$ and the subsequent values are regular. Thus the value $-1$ for $x_1$ introduces a singularity but the latter is confined, disappearing after five steps. We remark that this is only possible if at that iterate the mapping becomes indeterminate, which is the same as saying that the numerator and denominator at that iterate have a certain number of common factors. Hence, simplifications start appearing in the successive iterates once the singularity is confined and they have as a result that the growth, which would have been exponential (with a dynamical degree equal to $(1+\sqrt 5)/2$) in their absence, becomes polynomial, leading to a dynamical degree equal to 1. (Note that, contrary to the above, if we assume now that $z$ vanishes we obtain the following pattern $0,\infty,\infty,0,\bol,\infty,\bol,0$, where $\bol$ stands for a finite value. In fact pursuing the iterations we find that the pattern $\{\bol,\infty,\bol,0,\infty,\infty,0\}$ repeats indefinitely; it is what we call a cyclic pattern). 

While obtaining the degrees directly from the iterates of the mapping is straightforward, it becomes easily unmanageable since one has to perform simplifications of polynomials of growing degrees. The situation becomes still worse in the case of non-integrable mappings. We show this in the case of mapping (\ztri). With the same initial conditions 1 and $z$ we find the following succession of $x_n$ (where, again, $Q_n$ is a polynomial of degree $n$ in $z$):
$$x_2={z^2+1\over z}$$
$$x_3={Q_4(z)\over z^2(z^2+1)}$$
$$x_4={Q_8(z)\over z(z^2+1)^2Q_4(z)}$$
$$x_5={zQ_{18}(z)\over (z^2+1)Q_4(z)Q_8(z)}$$
$$x_6={(z^2+1)^2Q_{4}(z)Q_{38}(z)\over (z^2+1)Q_{4}^2(z)Q_{8}^2(z)Q_{18}(z)}$$
$$x_7={(z^2+1)Q_{4}^2(z)Q_{8}(z)Q_{80}(z)\over z(z^2+1)Q_{4}(z)Q_{8}^2(z)Q_{18}(z)Q_{38}^2(z)}$$
$$x_8={(z^2+1)^2Q_{4}^2(z)Q_{8}^2(z)Q_{18}(z)Q_{166}(z)\over z(z^2+1)^2Q_{4}^2(z)Q_{8}(z)Q_{18}^2(z)Q_{38}^2(z)Q_{80}(z)}$$
The corresponding degrees are 0, 1, 2, 4, 9, 19, 46, 98, 213, without taking into account any possible simplifications. After simplification, we find the sequence 0, 1, 2, 4, 9, 19, 40, 84, 175. As in the previous case, a singularity appears if $z$ is equal to $\pm i$. Then $x_2$ vanishes and the  successive values of $x_n$ are $\pm i,0,\infty,\infty^2,\infty,0,\mp i$. (The symbol $\infty^2$ is introduced here since the term $z^2+1$ vanishing in the denominator of $x_4$ has multiplicity 2). This is again a confined singularity. A cyclic one does also exist: assuming that $z$ vanishes, we obtain the sequence $\bol,0,\infty,\infty^2,\infty,0,\bol,\infty,\infty$ and, in fact, this pattern repeats indefinitely.
Note that while some simplifications do appear, due to the presence of a confined singularity, they do not suffice to curb the exponential growth and the ratio of the degrees of $x_8$ to $x_7$ is already 2.08, i.e. very close to the dynamical degree.

Thus, computing the degree growth of a given birational mapping by iterating some initial condition and obtaining the successive rational expressions is not a very efficient method, due to the simplifications involved. But it is precisely these simplifications which condition the integrability of the mapping so they cannot be ignored. Fortunately a much more efficient method for the obtention of the degrees does exist. 
\bigskip
3. {\scap Obtaining the degrees with the help of singularities}
\medskip
In the previous section we have presented an explicit calculation of the degrees of the successive iterates of two second-order birational mappings, starting from initial conditions $x_0=1$ and $x_1=z$. The choice of the value 1 for $x_0$ is not constraining, meaning that the value 1 is not special. In fact, the idea behind the method of Halburd, which we shall present in what follows, is that $x_0$ must be generic (in the sense that it does not satisfy any special relation). On the other hand, the value $z$ of $x_1$ can take values freely in the closure of {\tenmsb C}. Starting from such initial conditions and iterating up to $x_n(z)$, one can obtain the degree $d_n$ as the number of pre-images of $x_n(z)=w$ for some $w$. Halburd's method is based on the observation that if one chooses for $w$ a value that appears in the singularity pattern, the computation of the degree is greatly simplified. 

Let us illustrate this is the case of (\zdyo). The confined singularity pattern is $\{-1,0,\infty,\infty,0,-1\}$. Asking what are the contributions to the number of preimages of $-1$ we see that either a $-1$ appears spontaneously at some iteration, ``opening'' the singularity pattern, or it appears 5 iterations later ``closing'' the pattern. If we denote the number of spontaneous occurrences of the value $-1$ at step $n$ by $M_n$ we have for the degree
$$d\equiv d_n(-1)=M_n+M_{n-5}.\eqdef\ztes$$
 But the degree must be the same when one considers the occurrences, i.e. the number of pre-images of 0 or $\infty$. We remark that a 0 appears if it is preceded by a $-1$ one or four steps earlier. Similarly the occurence of an $\infty$ is linked to the existence of $-1$ two or three steps earlier. We find thus for the degrees the relations
$$d_n(0)=M_{n-1}+M_{n-4},\eqdef\zpen$$
$$d_n(\infty)=M_{n-2}+M_{n-3},\eqdef\zhex$$
which however are not complete. In fact, 0 and $\infty$ also appear in the cyclic pattern $\{\infty,\bol,0,\infty,\infty,0,\bol\}$, which has period 7, and thus the right-hand side of (\zpen) and (\zhex) must be complemented by period-7 functions, accounting for the presence of 0 and $\infty$ in the cyclic pattern, and obtained by the repetition of the strings $[0,1,0,1,1,0,1]$ and $[0,1,1,1,1,1,1]$ (where the first position corresponds to $n=0\ {\rm mod}\,7$). Equating the three expressions for the degree (\ztes), (\zpen) and (\zhex), one obtains linear equations for $M_n$. For instance, subtracting (\zpen) from (\zhex) we find $M_{n+3}-M_{n+2}-M_{n+1}+M_n=[0,1,0,0,0,1,0]$. The corresponding characteristic polynomial is just $(k+1)(k-1)^2$ and the solution one finds for $M_n$ is $M=n^2/14+[-2,2,-2,0,1,1,0]/7$. The degree of the successive iterates grows quadratically with $n$
$$d_n={n^2+\psi_7(n)\over7},\eqdef\zhep$$
where $\psi_7(n)$ is a period-7 function, obtained by the periodic repetition of the string  $[0,6,3,5,5,3,6]$.

Before proceeding to the presentation of the algorithmic method of computing the degrees that is the subject of our paper, let us deal with the non-integrable mapping (\ztri). As we saw the confined singularity pattern is $\{\pm i,0,\infty,\infty^2,\infty,0,\mp i\}$. Here there are two ways to enter the singularity: either by a $+i$ or by a $-i$. We denote by $I_n$ and $M_n$ the spontaneous occurrences of $+i$ and $-i$ respectively at iteration $n$. We remark that a $-i$ may appear either spontaneously or due to the presence of a $+i$ six iterations earlier and similarly for the appearance of $+i$. We have thus for the degrees 
$$d_n(i)=I_n+M_{n-6},\eqdaf\zoct$$
$$d_n(-i)=M_n+I_{n-6}.\eqno (\zoct \rm b)$$
The occurence of 0 is linked to the appearance of either a $+i$ or a $-i$ one or five steps earlier while infinity appears due to the presence of $\pm i$ two, three (with multiplicity 2) or four steps earlier. 
We obtain thus the expressions for the degrees
$$d_n(0)=I_{n-1}+I_{n-5}+M_{n-1}+M_{n-5},\eqdaf\zenn$$
$$d_n(\infty)=I_{n-2}+2I_{n-3}+I_{n-4}+M_{n-2}+2M_{n-3}+M_{n-4}.\eqno(\zenn \rm b)$$
As in the case of (\zdyo) one must add to the right-hand sides of (\zenn) periodic terms, with period 9, to account for the appearance of 0 and $\infty$ in the cyclic pattern $\{0,\infty,\infty^2,\infty,0,\bol,\infty,\infty,\bol,\cdots\}$. However since we are only interested in this non-integrable case in obtaining the dynamical degree, we can apply our ``express'' version  [\refdef\express] of Halburd's method which does away with these periodic contributions. Equating the expressions for the degrees we find a linear equation for the quantity $I_n+M_n$, the characteristic polynomial of which is $k^6-k^5-k^4-2k^3-k^2-k+1$ which factorises into $(k^2+1)(k^4-k^3-2k^2-k+1)$. The largest root of the latter is $\mu+\sqrt{\mu^2-1}$ with $\mu=(1+\sqrt{17})/4$
, precisely  the dynamical degree that we had obtained in [\redem], using the full-deautonomisation approach.

The method of Halburd, as we have presented it, leads, at least in the case of an integrable mapping, to a closed expression for the degrees of the iterates. This requires the solution of some linear system of equations, a task which, admittedly, does not present particular difficulties. In the case of non-integrable mappings however the same approach usually does not yield a useful handle for the computation of the degrees 
and the fact that the degrees grow exponentially precludes the use of the direct, brute force, approach, based on the simplifications of rational expressions. Fortunately, Halburd's method of working with the values that appear in a singularity pattern can be cast in a way that is both convenient for calculations and can be described as an algorithm allowing the computation of the degrees to very high orders. In  [\refdef\highorder] we outlined this method but without insisting on the fact that it can be cast into a simple algorithm. In what follows we start by explaining how the method works for the first few iterations and then proceed to its algorithmic description. We start with mapping (\zdyo) and remark that the string $\{\infty,\bol,0,\infty,\infty,0,\bol\}$ introduces in fact two possible ways to enter the cyclic pattern: $\{\bol,\infty,\bol,0,\infty,\infty,0\}$ and 
$\{\bol,0,\infty,\infty,0,\bol,\infty\}$. Since the degree is the same whether we compute it from the occurrences (i.e. the number of pre-images) of 0, $\infty$ or $-1$, we must balance the 0 and $\infty$ of the cyclic patterns by a $-1$ 
 for a confined pattern. This means that at the next iteration step an $\infty$ (from the cyclic pattern) and a 0 (from the confined pattern) are present. Again they must be balanced by a $-1$ 
 for a  new  confined pattern. So at the next iteration we have two 0 and two $\infty$ from the cyclic and confined patterns. They must be balanced by two $-1$ 
  for two  new  confined patterns. 
 And so on. This leads to the tableau below. We have adopted the following convention for the setup of the tableau. The abscissa corresponds to the iterations while on the ordinate we represent the values that appear in the singularity pattern. In each cell we give the singularities in the upper right corner while their multiplicity is given in the lower left corner. The degree for each iteration is obtained by adding the multiplicities corresponding to one of the values appearing in the singularity pattern. (Since the singularities are balanced any one of those leads to the same result). We obtain thus the tableau \figdef\one. 
\medskip
\centerline{\includegraphics[width=8 cm,keepaspectratio]{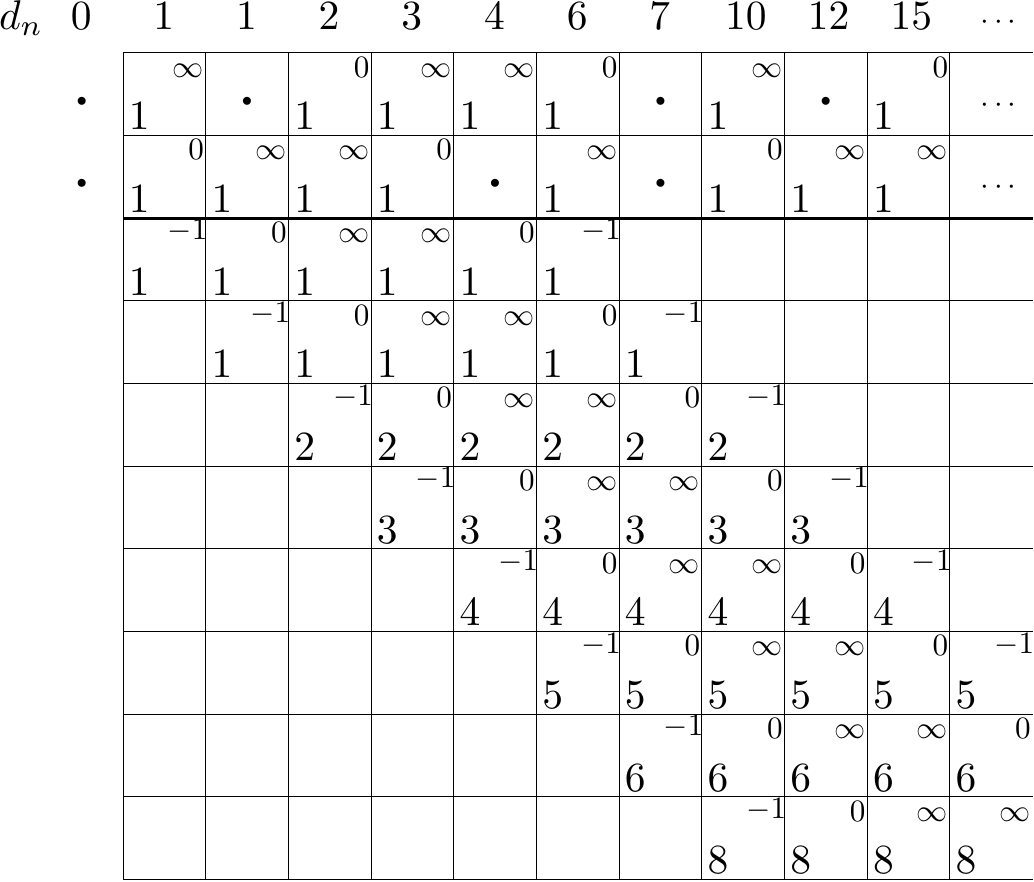}}
\smallskip{\bf Tableau \one}. {\sl Balancing of the singularities for the mapping (\zdyo). The degrees are given just above the tableau. The first two lines in the tableau correspond to the cyclic pattern (the dots on the left side represent the finite values preceding the two possible entries). In each cell we give the singularities in the upper right corner while their multiplicity is given in the lower left corner.}

\smallskip
We turn now to the non-integrable mapping (\ztri). Here again the string 
$\{0,\infty,\infty^2,\infty,0,\bol,\infty,\infty,\bol\}$ leads to two entry possibilities for the cyclic pattern: $\{\bol,\infty,\infty,\bol, 0,\infty,\infty^2,\infty,0\}$ and $\{\bol,0,\infty,\infty^2,\infty,0,\bol,\infty,\infty\}$. There are also two confined singularity patterns, corresponding to the sign choice in $\{\pm i,0,\infty,\infty^2,\infty,0,\mp i\}$. Just as in the previous case we set up a tableau by balancing the singularities at each iteration. Since the quantity $\infty^2$ appears in the singularity and in the cyclic patterns it must counted twice when it comes to obtaining the multiplicities of each singularity. We thus obtain the tableau \figdef\two:
\medskip
\centerline{\includegraphics[width=9 cm,keepaspectratio]{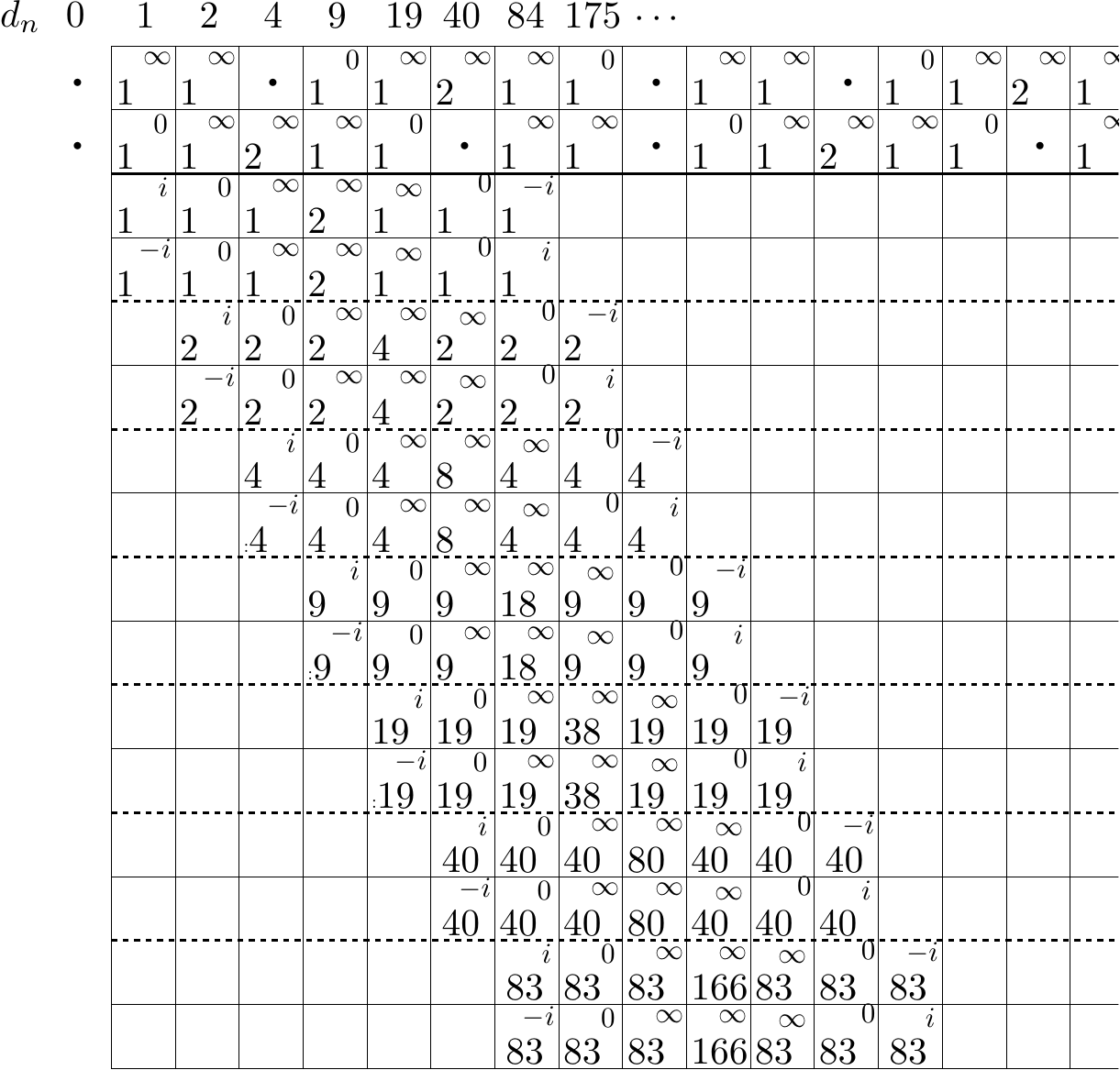}}
\smallskip{\bf Tableau \two}. {\sl Balancing the singularities for the mapping (\ztri). The degrees are given just above the tableau. The first two lines in the tableau corresponds to the cyclic pattern (the dots on the left side represent the finite values preceding the two possible entries).}

\smallskip
Balancing the singularities, as was done in the two tableaux above, can be cast into an algorithm, the calculations being performed automatically. We start by introducing variables for the preimages of the various quantities that appear in the (confined) singularity pattern. In the case of mapping (\zdyo) we have $M_n$ (already introduced in the previous paragraphs)  as well as  $Z_n$  and $F_n$,  for zero and  infinity respectively. Here we define objects with two indices $M_{n,m}$, $Z_{n,m}$, $F_{n,m}$ corresponding to the two dimensions of the tableau, where $n$ refers to the column index which starts at 1.

To start, we set their values equal to 0 for all $m$ and $n$. Then we define the contribution coming from the cyclic pattern by assigning values to $F$, $Z$ and $M$. If more than one cyclic patterns exists, this must be done for each of them. However when the cyclic patterns have the same periodicity, it is simpler to merge them into a single one combining the contributions. In the case of 
 mapping (\zdyo) for instance we have $F_{1,0}=F_{2,0}=F_{3,0}=F_{4,0}=F_{5,0}=F_{6,0}=1$, $F_{7,0}=0$ , $Z_{1,0}=Z_{3,0}=Z_{4,0}=Z_{6,0}=1$ and $Z_{2,0}=Z_{5,0}=Z_{7,0}=0$, combined with the periodicity condition $F_{n+7,0}=F_{n,0}$, $Z_{n+7,0}=Z_{n,0}$. And since $-1$ does not appear in the cyclic pattern we have  $M_{n,0}=0$ for all $n$. 

Next we introduce the initial condition for the confined singularity pattern. We have, always for  mapping (\zdyo),  $M_{1,1}=M_{6,1}=1$, $Z_{2,1}=Z_{5,1}=1$ and $F_{3,1}=F_{4,1}=1$. The condition for balancing the singularities at any iteration is
$M_{n,n}=\sum_{k=0}^{n-1}Z_{n,k}-M_{n,n-5}$, which is valid for $n>5$ (but it suffices to set $M_{n,k}=0$ for $k<0$ for the relation to be valid everywhere). The $M_{n,n-5}$ term corresponds to the value $-1$ that ``closes'' the singularity pattern and must be subtracted, lest we double-count the contribution of the singularity $-1$.
It goes without saying that the sum entering the expression for $M_{n,n}$ could have run over the $F$ instead of the $Z$. Once $M$ is obtained we compute the remaining quantities on the $n$-th line as
$F_{n+3,n}=F_{n+4,n}=M_{n,n}$, $Z_{n+2,n}=Z_{n+5,n}=M_{n,n}$ and $M_{n+6,n}=M_{n,n}$. We iterate this process up to some arbitrarily high $N$ and compute the degree as $d_n=\sum_{k=0}^NM_{n,k}$.

The computations based on the algorithm just described do not necessitate computer algebra: they can be carried out in any programming language using integer arithmetic. As a consequence the calculations are incredibly fast and allow one to obtain the succession of the degrees up to very high orders. We illustrate this by presenting the first 100 degrees for the mapping (\zdyo): 

0, 1, 1, 2, 3, 4, 6, 7, 10, 12, 15, 18, 21, 25, 28, 33, 37, 42, 47, 52, 58, 63, 70, 76, 83, 90, 97, 105, 112, 121, 129, 138, 147, 156, 166, 175, 186, 196, 207, 218, 229, 241, 252, 265, 277, 290, 303, 316, 330, 343, 358, 372, 387, 402, 417, 433, 448, 465, 481, 498, 515, 532, 550, 567, 586, 604, 623, 642, 661, 681, 700, 721, 741, 762, 783, 804, 826, 847, 870, 892, 915, 938, 961, 985, 1008, 1033, 1057, 1082, 1107, 1132, 1158, 1183, 1210, 1236, 1263, 1290, 1317, 1345, 1372, 1401.

The ease and power of our method become even more apparent in the case of the non-integrable mapping (\ztri). The computation of the degrees up to numbers with 32 digits takes up a negligible amount of time resulting to: 
 \smallskip
\noindent
\seqsplit{%
0,\ 1,\ 2,\ 4,\ 9,\ 19,\ 40,\ 84,\ 175,\ 364,\ 759,\ 1580,\ 3288,\ 6843,\ 14241,\ 29636,\ 61674,\ 128345,\ 267088,\ 555817,\ 1156666,\ 2407044,\ 5009105,\ 10424043,\ 21692632,\ 45142780,\ 93942983,\ 195497132,\ 406833247,\ 846627716,\ 1761848360,\ 3666439907,\ 7629931097,\ 15878031556,\ 33042485298,\ 68762039601,\ 143095110656,\ 297783643601,\ 619693419218,\ 1289593777476,\ 2683669148857,\ 5584766479427,\ 11622005135400,\ 24185613465636,\ 50330721067007,\ 104739186654252,\ 217964237118503,\ 453587718028380,\ 943924657852632,\ 1964325144373643,\ 4087797940988785,\ 8506785169560324,\ 17702781538058906,\ 36839824673794697,\ 76664374978484048,\ 159540020694572025,\ 332005813787275914,\ 690910405481109316,\ 1437797678771749121,\ 2992084282826671643,\ 6226584232064003288,\ 12957640071007986380,\ 26965095139190915479,\ 56114875230444219884,\ 116776121347770033935,\ 243013326876841402804,\ 505715349663634775080,\ 1052403249534643394739,\ 2190071154390984313769,\ 4557579676247064475524,\ 9484409884900041722722,\ 19737237142250511592801,\ 41073565433906675200000,\ 85474869927060675632801,\ 177874828052224495902882,\ 370160896198002010775684,\ 770311856795605003014249,\ 1603033607316772844835699,
3335943389053760365737000,\ 6942161564284909047646964,\ 14446770092913597620942415,
30064003003220403237137644,\ 62563761364278747160932439,\ 130196375899348242208503180,\ 
270941131538212542146563288,\ 563833641697967370487364443,\ 1173348519309461949828061761,\ 
 2441760558344260990740850756,\ 5081350107122939718737775434,\ 10574386101422956279560174265,\ 22005498354703634757948514128,\ 45793860106328226045065787337,\ 95297892810035512121785214426,\ 198316725275972642690305128964,\ 412700872647668258220992831025,\ 858838355903320829678322516043,\ 1787258933664594476688828092632,\ 3719319792842931751576160826780,\ 7739975143427773276411146697063,\ 16107035306874910426573973927212,\ 33519046452908794254283600055487}

Having presented the detail of our algorithmic method for the computation of the degree growth, we turn now to its application to some illustrative examples.
\bigskip
4. {\scap A collection of examples}
\medskip
In this section we are going to present the application of the algorithm we  introduced in the previous section to the computation of the degree growth of second- or higher-order mappings, integrable or non-integrable and possessing confined or unconfined singularities, to show that our approach can be applied in all cases studied and indeed leads to the correct value of the dynamical degree. 
\medskip
{\sl An integrable mapping (autonomous limit of discrete Painlev\'e I)}
\smallskip
We start with the mapping
$$x_{n+1}+x_n+x_{n-1}={a\over x_n}+1,\eqdef\zdek$$
depending on a non-zero constant $a$.
It possesses a confined singularity pattern $\{0,\infty,\infty,0\}$ as well as a cyclic one with the string $\{\bol,\infty,\infty\}$, repeated indefinitely. The equations for the degree can be obtained in a straightforward way. Introducing the quantity $Z_n$ for the number of spontaneous occurences of 0 at step $n$ we have for the degree $d_n(0)=Z_n+Z_{n-3}$. Similarly linking the degree to the occurences of infinity we find $d_n(\infty)=Z_{n-1}+Z_{n-2}$ to which one must add a function of pertiod 3 to account to the presence of infinity in the cyclic pattern. We could proceed to integrate the equation for $Z_n$ obtained by equating the two expressions for the degree, but instead we go on directly to apply the algorithm introduced in the previous section.
Starting from initial conditions, a generic value for $x_0$, and $x_1=z$, we compute the degrees in $z$ of the iterates using our algorithm. We obtain the following sequence for the first 100 terms:

0, 1, 2, 3, 6, 9, 12, 17, 22, 27, 34, 41, 48, 57, 66, 75, 86, 97, 108, 121, 134, 147, 162, 177, 192, 209, 226, 243, 262, 281, 300, 321, 342, 363, 386, 409, 432, 457, 482, 507, 534, 561, 588, 617, 646, 675, 706, 737, 768, 801, 834, 867, 902, 937, 972, 1009, 1046, 1083, 1122, 1161, 1200, 1241, 1282, 1323, 1366, 1409, 1452, 1497, 1542, 1587, 1634, 1681, 1728, 1777, 1826, 1875, 1926, 1977, 2028, 2081, 2134, 2187, 2242, 2297, 2352, 2409, 2466, 2523, 2582, 2641, 2700, 2761, 2822, 2883, 2946, 3009, 3072, 3137, 3202, 3267, 3334, 3401, 3468, 3537, 3606, 3675, 3746, 3817, 3888, 3961, 4034, 4107, 4182, 4257, 4332, 4409, 4486, 4563, 4642, 4721, 4800, 4881, 4962, 5043, 5126, 5209, 5292, 5377, 5462, 5547, 5634, 5721, 5808, 5897, 5986, 6075, 6166, 6257, 6348, 6441, 6534, 6627, 6722, 6817, 6912, 7009, 7106, 7203, 7302, 7401, 7500, 7601, 7702, 7803, 7906, 8009, 8112, 8217, 8322, 8427, 8534, 8641, 8748, 8857, 8966, 9075, 9186, 9297, 9408, 9521, 9634, 9747, 9862, 9977.

In this case it is easy to show that the sequence above can be represented by the expression
$$d_n={n^2+\psi_3(n)\over3},\eqdef\dena$$
where $\psi_3(n)$ is a periodic function obtained by the repetition of the string $[0,2,2]$. 
\medskip
{\sl A non-integrable mapping with confined singularities}
\smallskip
The mapping
$$x_{n+1}+x_{n-1}=x_n+{a\over 1-x_n^2},\eqdef\ddyo$$
with constant $a$, has two confined singularity patterns $\{\pm1,\infty,\infty,\mp1\}$ and a cyclic one, obtained from the repetitions of the string $\{\bol,\infty,\infty\}$. Here we introduce the quantities $U_n$ and $V_n$ for the number of spontaneous occurences of 1 and $-1$ at step $n$ and find for the degree $d_n(1)=U_n+V_{n-3}$, $d_n(-1)=V_n+U_{n-3}$ and $d_n(\infty)=U_{n-1}+U_{n-2}+V_{n-1}+V_{n-2}$ (to which one must add a function of pertiod 3 to account to the presence of infinity in the cyclic pattern).
Again we introduce initial conditions $x_0,x_1=z$ and compute the degree in $z$ based on the algorithm we described earlier.  We obtain the sequence:
 \smallskip
\noindent
\seqsplit{%
0,\ 1,\ 3,\ 8,\ 23,\ 61,\ 160,\ 421,\ 1103,\ 2888,\ 7563,\ 19801,\ 51840,\ 135721,\ 355323,\ 930248,\ 2435423,\ 6376021,\ 16692640,\ 43701901,\ 114413063,\ 299537288,\ 784198803,\ 2053059121,\ 5374978560,\ 14071876561,\ 36840651123,\ 96450076808,\ 252509579303,\ 661078661101,\ 1730726404000,\ 4531100550901,\ 11862575248703,\ 31056625195208,\ 81307300336923,\ 212865275815561,\ 557288527109760,\ 1459000305513721,\ 3819712389431403,\ 10000136862780488,\ 26180698198910063,\ 68541957733949701,\ 179445175002939040,\ 469793567274867421,\ 1229935526821663223,\ 3220013013190122248,\ 8430103512748703523,\ 22070297525055988321,\ 57780789062419261440,\ 151272069662201796001,\ 396035419924186126563,\ 1036834190110356583688,\ 2714467150406883624503,\ 7106567261110294289821,\ 18605234632923999244960,\ 48709136637661703445061,\ 127522175280061111090223,\ 333857389202521629825608,\ 874049992327503778386603,\ 2288292587779989705334201,\ 5990827771012465337616000,\ 15684190725257406307513801,\ 41061744404759753584925403,\ 107501042489021854447262408,\ 281441383062305809756861823,\ 736823106697895574823323061,\ 1929027937031380914713107360,\ 5050260704396247169315999021,\ 13221754176157360593234889703,\ 34615001824075834610388670088,\ 90623251296070143237931120563,\ 237254752064134595103404691601,\ 621141004896333642072282954240,\ 1626168262624866331113444171121,\ 4257363782978265351268049559123,\ 11145923086309929722690704506248,\ 29180405475951523816804063959623,\ 76395293341544641727721487372621},

where we have eliminated from the list all degrees longer than 32 digits. Computing the dynamical degree from the values for the degrees of the list we obtain of  2.618034, in perfect agreement with the exact value $(3+\sqrt 5)/2$ of the dynamical degree of 
(\ddyo).

\medskip
{\sl A mapping non-integrable due to bad deautonomisation}
\smallskip
We examine the mapping
$$x_{n+1}+x_{n-1}=1+{a_n\over x_n},\eqdef\dtri$$
where $a_n$ is a function of $n$. When $a_n$ is linear in $n$, the equation is a well-known integrable one: it is in fact a discrete analogue of Painlev\'e I  [\refdef\frompv]. But suppose we opt for a different $n$-dependence, taking for instance an $a_n$ quadratic in $n$. In this case the singularity confinement conditions are not satisfied and the singularity pattern which was confined as $\{0,\infty,\bol,\infty,0\}$ in the autonomous or $a_n$ linear in $n$ case, now becomes unconfined
$\{0,\infty,\bol,\infty,0,\infty,\bol,\infty,0,\infty,\bol,\infty,0,\cdots\}$ with the bloc $\{\infty,\bol,\infty,0,\}$ repeating indefinitely. The cyclic pattern $\{\bol,\infty,\cdots\}$, present in the integrable case, still exists. Let us show first how to deal with the unconfined pattern when it comes to balancing the singularities. Since the unconfined pattern is preceded by a semi-infinite sequence of regular values, just like a confined pattern, the way to deal with it is exactly the same as for the latter. The only difference is that at every iteration step we introduce a semi-infinite sequence of singularities that must be balanced. Tableau \figdef\three\ below shows the first few iterations:
\medskip
\centerline{\includegraphics[width=10 cm,keepaspectratio]{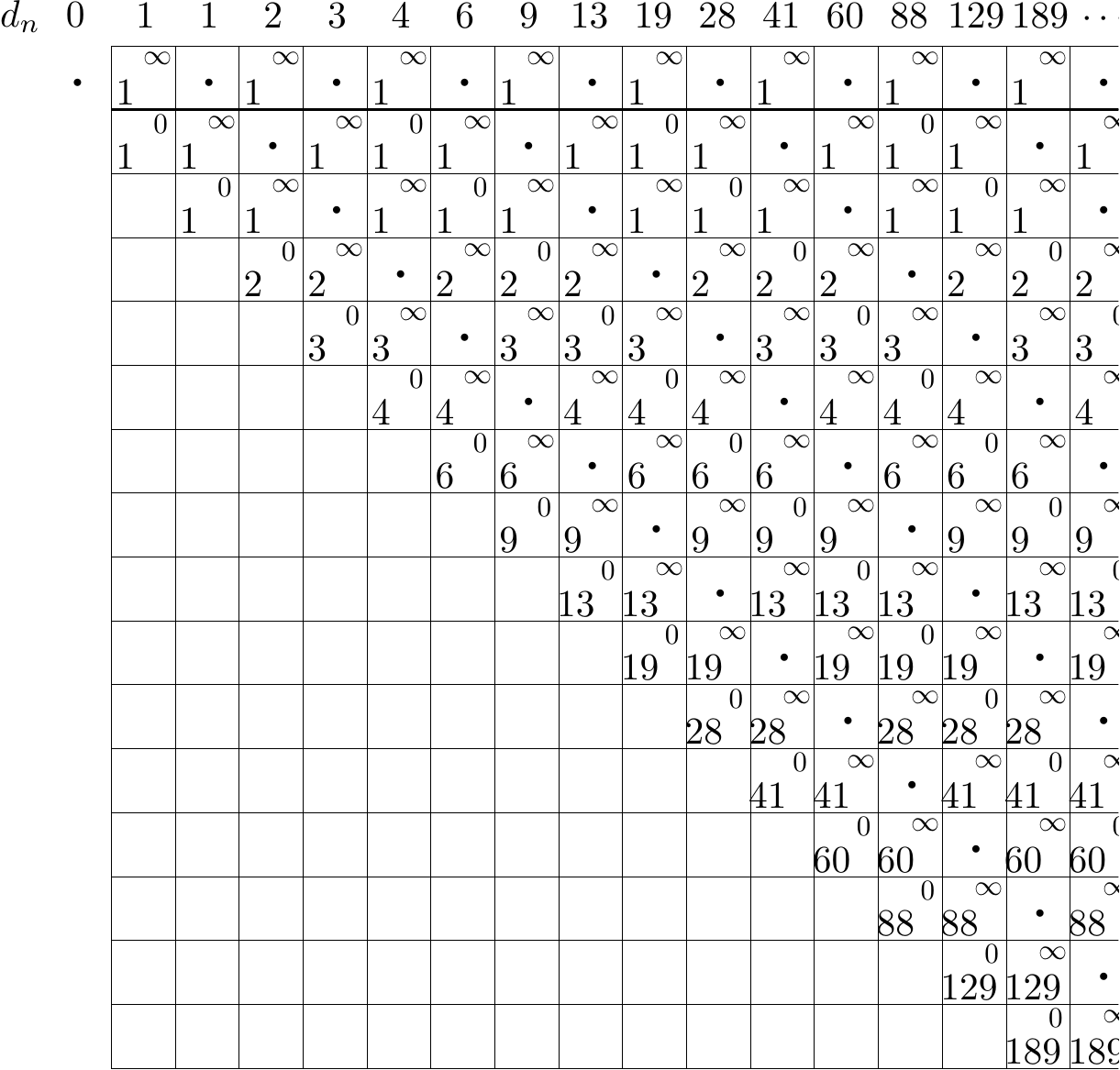}}
\smallskip{\bf Tableau \three}. {\sl Balancing the singularities for the mapping (\dtri). The degrees are given just above the tableau. The first line in the tableau corresponds to the cyclic pattern (the dot on the left side represents the finite value preceding the entry).}

\smallskip
Already inspecting the degrees presented in tableau 3, we see that the dynamical degree must be larger than 1, the ratio of 189 to 129 being equal to 1.465. In fact it is possible to implement the Diophantine approximation and obtain a better estimate of the dynamical degree, $\lambda=1.4655$ (where we are giving the converged decimals). Using Halburd's method the equations for the degree can be obtained formally if one introduces $Z_n$ for the number of spontaneous occurences of 0 at step $n$. We find $d_n(0)=Z_n+Z_{n-4}+Z_{n-8}+\cdots$ and $d_n(\infty)=Z_{n-1}+Z_{n-3}+Z_{n-5}+Z_{n-7}+\cdots$ and $(1+(-1)^n)/2$ must be added to the latter to account for the presence of infinity in the cyclic pattern. Given that we must now deal with infinite sums of possible contributions, it is far simpler (and faster) to implement our algorithm and obtain the sequence of degrees. We find
 \smallskip
 \noindent
\seqsplit{%
0,\ 1,\ 1,\ 2,\ 3,\ 4,\ 6,\ 9,\ 13,\ 19,\ 28,\ 41,\ 60,\ 88,\ 129,\ 189,\ 277,\ 406,\ 595,\ 872,\ 1278,\ 1873,\ 2745,\ 4023,\ 5896,\ 8641,\ 12664,\ 18560,\ 27201,\ 39865,\ 58425,\ 85626,\ 125491,\ 183916,\ 269542,\ 395033,\ 578949,\ 848491,\ 1243524,\ 1822473,\ 2670964,\ 3914488,\ 5736961,\ 8407925,\ 12322413,\ 18059374,\ 26467299,\ 38789712,\ 56849086,\ 83316385,\ 122106097,\ 178955183,\ 262271568,\ 384377665,\ 563332848,\ 825604416,\ 1209982081,\ 1773314929,\ 2598919345,\ 3808901426,\ 5582216355,\ 8181135700,\ 11990037126,\ 17572253481,\ 25753389181,\ 37743426307,\ 55315679788,\ 81069068969,\ 118812495276,\ 174128175064,\ 255197244033,\ 374009739309,\ 548137914373,\ 803335158406,\ 1177344897715,\ 1725482812088,\ 2528817970494,\ 3706162868209,\ 5431645680297,\ 7960463650791,\ 11666626519000,\ 17098272199297,\ 25058735850088,\ 36725362369088,\ 53823634568385,\ 78882370418473,\ 115607732787561,\ 169431367355946,\ 248313737774419,\ 363921470561980,\ 533352837917926,\ 781666575692345,\ 1145588046254325,\ 1678940884172251,\ 2460607459864596,\ 3606195506118921,\ 5285136390291172,\ 7745743850155768,\ 11351939356274689,\ 16637075746565861,\ 24382819596721629,\ 35734758952996318,}
 
and computing the dynamical degree from the ratio of the two last degrees we find the value $\lambda=1.465571232$. It turns out that, thanks to the methods we developed in  [\refdef\noncon], it is possible to provide an exact expression for the dynamical degree in a very simple way. The unconfined pattern can be considered as resulting form a confinement condition that has been infinitely delayed. Thus the characteristic equation which was $1-\lambda-\lambda^3+\lambda^4=0$ in the integrable case becomes now $1-\lambda-\lambda^3+\lambda^4-\lambda^5-\lambda^7+\lambda^8-\lambda^9-\lambda^{11}+\lambda^{12}-\cdots=0$.
We remark that the geometric series that appears can be easily summed up to some order $m$. We find thus for the characteristic equation
$$1-\left({\lambda^{4m}-1\over\lambda^4-1}\right)\lambda(1+\lambda^2-\lambda^3)=0\eqdef\dtes$$
and since $\lambda$ is larger than 1, at the limit when $m$ goes to infinity (i.e. an infinitely delayed confinement) the dynamical degree is given by the (supergolden ratio) equation $\lambda^3-\lambda^2-1=0$. The latter has one real root, its value coinciding with the dynamical degree previously computed. 

Having shown how our method works in the case of 
second order mappings, integrable or not, we now proceed to apply it to higher-order ones. A caveat is in order at this point. As stressed in a recent work of ours [\highorder], the way to proceed to the calculation of the degree growth and the dynamical degree follows the spirit of that employed for second-order mappings. Namely, for a mapping of order $N$ we start with an initial condition where $x_0, \cdots, x_{N-2}$ are generic and $x_{N-1}$ is free to take {\sl any} value including special ones. (`Special'  in this context is to be understood not only as a value appearing in a singularity pattern but, in fact, any value which, combined with the previous generic ones, may lead to a singularity at some later iteration). And we proceed to compute the degree of the successive iterates in $x_{N-1}$. While there is no guarantee that this suffices (as it did for second order mappings) in order to obtain the right value of the dynamical degree, it turned out that, in all cases examined, the value of the dynamical degree obtained was the expected one. The examples that we shall present in the following confirm this result, justifying further our approach.
\medskip
{\sl A coupling of two linearisable mappings}
\smallskip
In  [\refdef\thrid] we presented the couplings of various mappings with linear or linearisable ones. In this subsection we consider one of those mappings, more precisely the coupling of what we dubbed a ``third-kind'' linearisable mapping [\refdef\thirdkind]
$$x_{n+1}x_{n-1}=x_{n}^2-1,\eqdaf\dpen$$
to a ``Gambier''-type mapping  [\refdef\gambier], through
$$x_n=z_{n+1}-z_n+{1\over z_n}-{1\over z_{n-1}}.\eqno(\dpen{\rm b})$$
The resulting fourth-order mapping has two singularity patterns, a confining one $ \{ 0,\infty\}$ and an anticonfining one $ \{\cdots, 0^3,0^2,0,\bol,\bol,\bol,\bol,\bol, \infty,\infty^2,\infty^3,\infty^4\cdots\}$. The degree growth was obtained in [\thrid] by direct computation, leading to the  sequence $d_n$= 0, 0, 0, 1, 3, 7, 13, 21, 31, 43, 57, 73, 91, 111, 133, 157$\cdots$, which, for $n>2$, can be represented by the expression $d_n=(n-2)(n-3)+1$. What is interesting in the present case is the anticonfined pattern. As we have explained in  [\refdef\anticon], the presence of an anticonfined pattern is not incompatible with integrability unless the growth of the multiplicities of the values that appear in this pattern is exponential. This is not the case here and thus we expect the coupled system to be integrable. In order to compute its degree growth we proceed, as in the previous cases,
to the balancing of singularities. However the situation is now complicated by the fact that instead of just three free values in the anticonfined pattern (as expected given its order) here we have apparently five. And since one can put the value $\infty$ to any position compatible with the number of consecutive `free' values one would have to balance the following singularity patterns (in the positive $n$ direction): $\{\bol,\bol,\bol,\infty,\infty^2,\infty^3,\cdots\}$, $\{\bol,\bol,\bol,\bol, \infty,\infty^2,\infty^3,\cdots\}$ and $\{\bol,\bol,\bol,\bol,\bol, \infty,\infty^2,\infty^3,\cdots\}$. However it turns out that the pattern of the middle is not acceptable: it exists only under a constraint between $z_0, z_1$ and $z_2$. We are thus left with the remaining two patterns and balancing the singularities for the first few iterations leads to the tableau \figdef\four\ below:  
\medskip
\centerline{\includegraphics[width=9 cm,keepaspectratio]{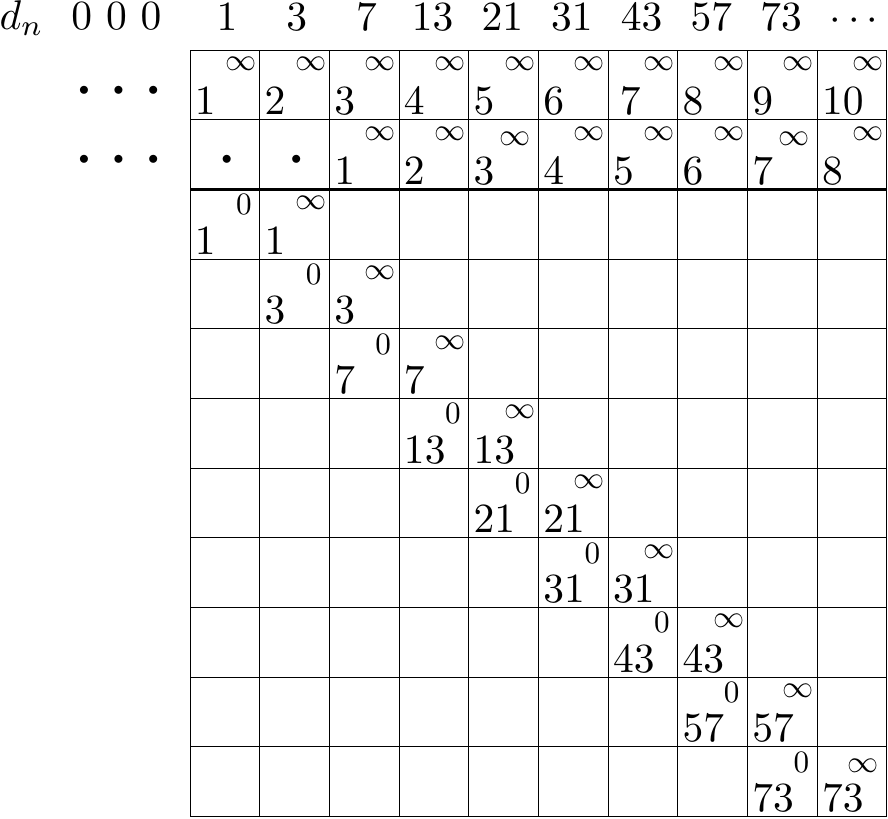}}
\smallskip{\bf Tableau \four}. {\sl Balancing the singularities for the mapping (\dpen). The degrees are given just above the tableau. The first two lines in the tableau correspond to the anticonfined pattern (the dots on the left side represent the finite values preceding the two acceptable entries).}
\smallskip
Introducing the quantity $Z_n$ for the number of spontaneous occurences of 0 at step $n$ we have for the degree $d_n(0)=Z_n$ and $d_n(\infty)=Z_{n-1}+2n-6$ the latter term coming for the contributions $n-2$ and $n-4$ of the anticonfined pattern.
Implementing our algorithm for the computation of the degrees is elementary resulting to the following sequence for the first hundred degrees:

0,\ 0,\ 0,\ 1,\ 3,\ 7,\ 13,\ 21,\ 31,\ 43,\ 57,\ 73,\ 91,\ 111,\ 133,\ 157,\ 183,\ 211,\ 241,\ 273,\ 307,\ 343,\ 381,\ 421,\ 463,\ 507,\ 553,\ 601,\ 651,\ 703,\ 757,\ 813,\ 871,\ 931,\ 993,\ 1057,\ 1123,\ 1191,\ 1261,\ 1333,\ 1407,\ 1483,\ 1561,\ 1641,\ 1723,\ 1807,\ 1893,\ 1981,\ 2071,\ 2163,\ 2257,\ 2353,\ 2451,\ 2551,\ 2653,\ 2757,\ 2863,\ 2971,\ 3081,\ 3193,\ 3307,\ 3423,\ 3541,\ 3661,\ 3783,\ 3907,\ 4033,\ 4161,\ 4291,\ 4423,\ 4557,\ 4693,\ 4831,\ 4971,\ 5113,\ 5257,\ 5403,\ 5551,\ 5701,\ 5853,\ 6007,\ 6163,\ 6321,\ 6481,\ 6643,\ 6807,\ 6973,\ 7141,\ 7311,\ 7483,\ 7657,\ 7833,\ 8011,\ 8191,\ 8373,\ 8557,\ 8743,\ 8931,\ 9121,\ 9313,\ 9507,\ 9703,\ 9901.
\medskip
{\sl Reductions of the Bogoyavlensky lattice}
\smallskip
The discrete form of the Bogoyavlensky system was obtained by Suris [\refdef\suris] and independently by Papageorgiou and Nijhoff  [\refdef\vasfrank]. One of the authors gave in  [\refdef\stef] the ``travelling wave'' reductions of these lattices. In what follows we shall work with one of the latter, namely
$$ {1-\prod_{k=0}^N x_{n+1+k}\over 1-\prod_{k=0}^N x_{n-k}}={x_{n+1}\over x_n}.\eqdef\dhex$$
Equation (\dtes) can be integrated once, leading to
$$x_n+\sum_{m=1}^N\prod_{k=0}^{N+1} x_{n+m-k}=C.\eqdef\dhep$$
The $N=1$ case gives, up to a rescaling, the mapping (\zdyo). For $N=2, 3, 4$ we find the mappings
$$x_{n+1}x_{n-1}(x_{n+2}+x_{n-2})={C\over x_n}-1,\eqdaf\doct$$
$$x_{n+1}x_{n-1}(x_{n+3}x_{n+2}+x_{n+2}x_{n-2}+x_{n-2}x_{n-3})={C\over x_n}-1.\eqno(\doct{\rm b})$$
$$x_{n+1}x_{n-1}\big(x_{n-2}x_{n-3}(x_{n+2}+x_{n-4})+x_{n+2}x_{n+3}(x_{n+4}+x_{n-2})\big)={C\over x_n}-1.\eqno(\doct{\rm c})$$
The integrated form (\dhep) has, for $N>1$, a confined singularity pattern 
$\{0,\infty,(N-1\ {\rm finite\ values}\ \bol),\infty,0\}$.
A cyclic pattern also exists,  but its details depend on $N$. For $N=2$ the cyclic pattern is obtained from the repetition of the string $\{\infty,\bol,\bol,\bol\}$. For $N=3$ the basic string is $\{\infty,\bol,0,\infty,\bol,\bol,\infty,0,\bol,\infty,\bol,\bol,\bol,\bol,\bol\}$, of length 15, while for $N=4$ we have a string of length 12, $\{\infty,\bol,0,\bol,\infty,\bol,\bol,\bol,\bol,\bol,\bol,\bol\}$. In fact, the length of the basic string of the cyclic patterns is equal to $N(N+2)$ for odd $N>1$, and $N(N+2)/2$ for even $N$. We focus now on the case $N=4$, a mapping of order 8. Introducing $Z_n$ for the number of spontaneous occurrences of the value 0 at step we find that the contribution of the preimages of 0 to the degree from the confined pattern is 
$$d_n(0)=Z_n+Z_{n-6}\eqdaf\denn$$
to which we must add a period-12 contribution coming form the cyclic pattern. Similarly for the contribution of the preimages of infinity we have
$$d_n(\infty)=Z_{n-1}+Z_{n-5}\eqno(\denn{\rm b})$$
and again there is a (different) period-12 contribution due to the cyclic pattern. Equating the two expressions for the degree we find the characteristic polynomial $(\lambda^5-1)(\lambda-1)$, which signals the presence of a period-5 term in the expression of $d_n$. And since there is a period-12 contribution coming from the cyclic pattern we expect the degree to include a period-60 term. Performing the calculation of the degrees in the standard way, through the simplification of the successive rational expressions, is prohibitively long. But this is not a real obstacle. Using the algorithm we introduced in the previous section, we compute a large number of degrees. A hundred of them are displayed below (but the calculation of thousands is literally instantaneous): 

0,\ 0,\ 0,\ 0,\ 0,\ 0,\ 0,\ 1,\ 1,\ 1,\ 0,\ 1,\ 2,\ 3,\ 2,\ 1,\ 2,\ 4,\ 5,\ 5,\ 4,\ 5,\ 6,\ 8,\ 8,\ 8,\ 8,\ 9,\ 11,\ 12,\ 12,\ 13,\ 14,\ 16,\ 16,\ 17,\ 18,\ 20,\ 21,\ 21,\ 22,\ 24,\ 26,\ 28,\ 28,\ 29,\ 30,\ 33,\ 35,\ 36,\ 36,\ 37,\ 40,\ 43,\ 44,\ 45,\ 46,\ 49,\ 51,\ 53,\ 54,\ 56,\ 58,\ 60,\ 62,\ 64,\ 66,\ 69,\ 71,\ 73,\ 74,\ 77,\ 80,\ 83,\ 84,\ 85,\ 88,\ 92,\ 95,\ 97,\ 98,\ 101,\ 104,\ 108,\ 110,\ 112,\ 114,\ 117,\ 121,\ 124,\ 126,\ 129,\ 132,\ 136,\ 138,\ 141,\ 144,\ 148,\ 151,\ 153,\ 156,\ 160,\ 164,\ 168,\ 170,\ 173,\ 176, $\dots$

Once the degrees are obtained we proceed to fit them with a quadratic-plus-periodic term. The result is
$$d_n={n^2-8n+1-\psi_{60}(n)\over60}$$
which reproduces perfectly the degrees  for $n>0$ and where $\psi_{60}(n)$ is a period-60 function obtained by the repetitions of the string
[0, 53, 48, 45, 44, 45, 48, 53, 0, 9, 20, 93, 48, 5, -36, 45, 128, 93, 0, -31, 0, 93, 68, 45, -36, 5, 48, 93, 80, 9, 0, 53, 48, 45, -16, 45, 48, 53, 0, 9, 80, 93, 48, 5, -36, 45, 68, 93, 0, -31, 0, 93, 128, 45, -36, 5, 48, 93, 20, 9].
\medskip
{\sl A mapping proposed by Viallet}
\smallskip
In [\refdef\viallet] Viallet introduced the fourth-order mapping
$$x_{n+2}x_{n-2}={x_{n+1}x_{n-1}\over x_{n}(1-x_{n})}\eqdef\ddek$$
and studied its properties. 
The mapping is non-integrable but, as we showed in [\highorder], it has a confined singularities with pattern
$\{1,\bol,\infty,\infty,0,0^2,0,0,0,\infty,\infty^2,\bol,0^2,0,0,0^2,\bol,\infty^2,\infty,0,0,0,0^2,0,\infty,\infty,\bol,1\}.$
A cyclic pattern does also exist,

$\{\bol,\bol,\bol,0,0,\bol,\bol,\bol,\infty,\infty,0,0^2,0,0,0,\infty,\infty^2,\bol,0^2,0,0,0^2,\bol,\infty^2,\infty,0,0,0,0^2,0,\infty,\infty,\cdots\}.$

Viallet calculated the dynamical degree of the mapping and obtained a value of  $\lambda=1.400618098$. In [\highorder] we confirmed his result, first by performing a calculation using the Diophantine method, secondly by using the express method for the calculation of the dynamical degree and thirdly by applying the full-deautonomisation approach. 

We are not going to give any details concerning those calculations here: they can be found in the aforementioned reference. Instead we shall show how the algorithmic method developed in this paper allows to calculate the degree of the successive iterates of the mapping up to very high order, giving their exact numerical values. Given the length of the singularity patterns we cannot present a tableau as for the other cases, but suffice it to say that because the two strings $[\bol,\bol,\bol,0]$ and $[\bol,\bol,\bol,\infty]$ that appear in the cyclic pattern, there are two entry possibilities and thus, in order to establish the balance of singularities, we must consider the cyclic pattern twice, once for each distinct entry. Calculating the degrees of the successive iterates up to numbers with 32 digits we find the sequence:

\noindent
\seqsplit{%
0,\ 0,\ 0,\ 1,\ 1,\ 1,\ 2,\ 2,\ 4,\ 5,\ 7,\ 11,\ 13,\ 21,\ 27,\ 39,\ 56,\ 74,\ 110,\ 148,\ 211,\ 296,\ 409,\ 583,\ 805,\ 1137,\ 1589,\ 2220,\ 3125,\ 4355,\ 6122,\ 8561,\ 11988,\ 16812,\ 23510,\ 32973,\ 46147,\ 64646,\ 90567,\ 126792,\ 177668,\ 248766,\ 348474,\ 488089,\ 683547,\ 957525,\ 1340969,\ 1878312,\ 2630758,\ 3684603,\ 5160931,\ 7228205,\ 10124236,\ 14180020,\ 19860756,\ 27817615,\ 38961372,\ 54570561,\ 76432069,\ 107052279,\ 149939657,\ 210007460,\ 294141267,\ 411978590,\ 577025261,\ 808192154,\ 1131967568,\ 1585455968,\ 2220616352,\ 3110237027,\ 4356253777,\ 6101446832,\ 8545799460,\ 11969397802,\ 16764558686,\ 23480742219,\ 32887552057,\ 46062904113,\ 64516531156,\ 90363028131,\ 126564087071,\ 177267952700,\ 248284706512,\ 347752044329,\ 487067819683,\ 682195991057,\ 955496059093,\ 1338285074905,\ 1874426284426,\ 2625355399233,\ 3677120262371,\ 5150241209020,\ 7213521042247,\ 10103388111876,\ 14150988276811,\ 19820130246014,\ 27760433178461,\ 38881765102819,\ 54458503894687,\ 76275566209288,\ 106832938422914,\ 149632147145442,\ 209577493323421,\ 293538030185392,\ 411134677688492,\ 575842670331675,\ 806535666047600,\ 1129648450708241,\ 1582206064969660,\ 2216066449995075,\ 3103862776940992,\ 4347326380558854,\ 6088944007905416,\ 8528285177786624,\ 11944870568143731,\ 16730201900482150,\ 23432623571899094,\ 32820156666958892,\ 45968505420209104,\ 64384320647061992,\ 90177844750615639,\ 126304721431861067,\ 176904678748378423,\ 247775894745546074,\ 347039402524002567,\ 486069668029109121,\ 680797974119525108,\ 953537963896201536,\ 1335542529734901571,\ 1870585038311532699,\ 2619975259233867999,\ 3669584765411676264,\ 5139686836002157877,\ 7198738402553836452,\ 10082683292172993102,\ 14121988699333644965,\ 19779512957514883590,\ 27703543825590987029,\ 38802084871609573672,\ 54346902326352142398,\ 76119254989614038154,\ 106614006174251491940,\ 149325506589293239414,\ 209148007079909727440,\ 292936483957867009444,\ 410292141109513412654,\ 574662598464288635564,\ 804882835875694547385,\ 1127333467009268246062,\ 1578963656811522231183,\ 2211525074426955280631,\ 3097502044280978247274,\ 4338417423013415995101,\ 6076465961033910313538,\ 8510808199261987730729,\ 11920391995794362546576,\ 16695916769187586022690,\ 23384603196101909963173,\ 32752898460079439467899,\ 45874302357847838179820,\ 64252378133319645626005,\ 89993043677989226761213,\ 126045885704402986860053,\ 176542148744917218261287,\ 247268128660416604553159,\ 346328216156283003233114,\ 485073567530880410120385,\ 679402817733327888901663,\ 951583882613022567489241,\ 1332805608122009518427854,\ 1866751656368554103347084,\ 2614606155105353528248079,\ 3662064700996912975961005,\ 5129154097683684709104535,\ 7183986058636136905461982,\ 10062020892292003555832968,\ 14093048568101168311547683,\ 19738978885941926307047355,\ 27646771071346539539732415,\ 38722567924514505893561355,\ 54235529451129113043360373,\ 75963264124900443401373673,\ 106395522545954499195852156,\ 149019494465299846283504013,\ 208719400960711821725749125,\ 292336170470250123359369254,\ 409451331174037516919182256,\ 573484944851367455523562861,\ 803233392911801831007474170,\ 1125023227341774573374324509,\ 1575727893321634484608393739,\ 2206993005520891540594895571,\ 3091154346548028114128212281,\ 4329526722685536031350771505,\ 6064013485247335678192491024,\ 8493367036305597799567522659,\ 11895963587300670007217826259,\ 16661701898136554995376614133,\ 23336681228473790510107047926,\ 32685778084909149303229895079,\ 45780292345613769694216481407,\ 64120706008754881456103081748,\ 89808621317270719787789006990.}

It goes without saying that the ratio of $d_n/d_{n-1}$ reproduces the value of the dynamical degree once $n$ is sufficiently  large.

\medskip
{\sl A rational but not birational mapping}
\smallskip
In  [\refdef\lattice] some of the present authors introduced, among others, the higher-order non-integrable mapping
$$x_{n+N}+x_{n-1}={1\over x_{n+N-1}^{2m}}+{1\over x_n^{2m}}\eqdef\vena$$
where $N>1$ and $m>1$. As shown in [\highorder] (\vena) has a confined singularity pattern
$\{0,\infty^{2m},(N-2 {\rm\ finite\ values\ \bol}),\infty^{2m},0\},$
as well as a cyclic one
$\{(N-1{\rm\ finite\ values\ \bol}), \infty)\}$.
Here we shall study a variant of that mapping
$$x_{n+N}+x_{n+1}={1\over x_{n+N-1}^m}+{1\over x_n^m},\eqdef\vdyo$$
where $N>1$ and $m>1$. This is {\sl not} a birational mapping: only the forward evolution is described in a rational way. It is elementary to verify the existence of a confined singularity $\{0,\infty^m,(N-2\ {\rm finite\,values})\}$.  The appearance of the finite values in the confined singularity pattern may appear strange at first sight. However as we have regularly explained a singularity is confined when all indeterminacies are lifted and one recovers all the degrees of freedom of the mapping. In the present case one needs these extra $N-2$ steps in order to recover all the missing degrees of freedom. 

But another singularity does also exist. Starting with $N-1$ finite values and assuming that the $N$th one is infinity one finds that the infinity does not disappear in the subsequent iterations. However this singularity is not unconfined but rather an anticonfined one. 

Let us start with the $N=2$ case with finite $x_0$ and assume that $x_1$ behaves like $1/\epsilon$ (and thus $x_1\to\infty$ when  $\epsilon\to0$). From the expression (\vdyo) of the mapping we have $x_1+x_0-x_0^{-m}-x_{-1}^{-m}=0$. This means that if $x_1$ and $x_{-1}$ are related through $x_1x_{-1}^{m}-1=0$ (up to sub-dominant terms) it is possible to have the $1/\epsilon$ behaviour we assume for $x_1$. Using the instances of  (\vdyo) 
for negative indices we find that the $x_{-n}$ are related to $x_1$ by $x_1x_{-n}^{m}+(-1)^n=0$ (always up to sub-dominant terms). On the other hand, iterating (\vdyo) forwards we find that if $x_1$ diverges then all the  subsequent values of $x$ for positive indices are infinite. This solution is not 
what we call unconfined: the values for negative $n$ are not regular either. In fact, they all vanish like $\epsilon^{1/m}$ . Only $x_0$ is finite. This solution is 
 anticonfined, of the form 
 $$ \{\dots,0^{1/m},0^{1/m},0^{1/m},x_0,\infty,\infty,\infty,\dots\},$$
where $x_0$ is the only free value.

In the case $N=3$ is we start with $x_0, x_1$ finite and assume that  $x_2$ goes to infinity like $1/\epsilon$ when $\epsilon\to 0$ we find by iterating a sequence  of infinities alternating with finite values. The appearance of $\infty$ with period 2 leads to a periodic term in the expression for the degree.
From  (\vdyo)  we have $x_2+x_0-x_1^{-m}-x_{-1}^{-m}=0$ and thus $x_1$ and $x_{-1}$ are related through $x_2x_{-1}^{m}-1=0$ (up to sub-dominant terms), i.e. a situation similar to that encountered in the $N=2$ case. Similarly we find that  $x_{-2}$ obeys the relation $x_1+x_{-1}-x_0^{-m}-x_{-2}^{-m}=0$ and since  $x_0, x_1$ are finite and $x_{-1}$ is not divergent, $x_{-2}$ has a finite value. Going one step backwards we find for $x_{-3}$ the relation $x_2-x_{-2}-x_1^{-m}+x_{-3}^{-m}=0$ and, neglecting sub-dominant terms, we find that $x_2$ and $x_{-3}$ must satisfy the relation $x_2x_{-3}^{m}+1=0$. The pattern now becomes clear: the $x$ with negative odd indices are such that they must compensate the divergence of $x_2$ through a relation  $x_2x_{1-2k}^{m}+(-1)^k=0$, for $k>0$. Again, an anticonfined pattern is present here of the form  
$$ \{\dots,0^{1/m},\bol,0^{1/m},\bol, 0^{1/m},x_0,x_1,\infty,\bol,\infty,\bol,\infty,\dots\}.$$
We have checked several values for $N$ and we expect the general singularity pattern to be 
$$\{\cdots,0^{1/m}, (N-2\ {\rm finite\,values}\ \bol), 0^{1/m}, (N-2\ {\rm finite\,values}\ \bol), x_0, x_1,\cdots, x_{N-2}, \infty, (N-2\ {\rm finite\,values}\ \bol), \infty,\cdots\}.$$

We can now apply Halburd's method used in the previous examples to compute the degree growth of the mapping. For positive $n$ let us call $Z_n$ the number of occurences of the value 0 at step $n>0$. The only appearance of a zero value is in the confined singularity pattern and the degree $d_n(0)$ is equal to $Z_n$.  For $n>1$, the infinite values for $x_n$ come either from the anticonfined singularity, which provides one such value for $n=1\ {\rm mod}\,(N-1)$ and $m$ values for each instance of the confined singularity, starting at $n-1$, namely $Z_{n-1}$. We have thus $d_{n}(\infty)=mZ_{n-1}+\psi_{N-1}(n)$ where $\psi_{N-1}(n)$ is a period-$(N-1)$ function.  The periodic function can be written in terms of the $(N-1)$th root of unity $\omega$ ($\omega^{N-1}=1$, with $\omega\ne1$) as $\psi_{N-1}(n)=\sum_{k=0}^{N-2} \omega^{k(n-1)}/(N-1)$.
(The case $N=2$ is special since there is an entry for infinity at every $n$ and thus $d_{n}(\infty)=mZ_{n-1}+1$).
 Finally we have for the degree the recursion 
$d_{n}=md_{n-1}+\psi_{N-1}(n)$ leading to a dynamical degree equal to $m$ (plus a subdominant term coming from the periodic function).

What is interesting with mapping (\vdyo) is that it contains 
a family of integrable sub-cases.
Indeed, for $m=1$ the mapping 
$$x_{n+N}+x_{n+1}={1\over x_{n+N-1}}+{1\over x_n},\eqdef\vtri$$
belongs to the Gambier family of linearisable mappings. It possesses a confined singularity $\{0,\infty,(N-2\ {\rm finite\,values})\}$. An anticonfined singularity is also present. It has the form 
$$\{\cdots,0, (N-2\ {\rm finite\,values}\ \bol), 0, (N-2\ {\rm finite\,values}\ \bol), x_0, x_1,\cdots, x_{N-2}, \infty, (N-2\ {\rm finite\,values}\ \bol), \infty,\cdots\}$$
which has the same structure as the singularity pattern found in the non-integrable, $m>1$, case.  And, as expected, the degree of the iterates is given by the recursion $d_n=d_{n-1}+\psi_{N-1}(n)$ with the periodic function $\psi$ defined in the previous paragraphs.  

Before concluding this section it is interesting to give the integration of (\vtri). We start by rewriting it as $x_{n+N}\,-1/x_{n+N-1}=-(x_{n+1}-1/x_n)$ and introduce $y_n=x_{n+1}-1/x_n$. The mapping becomes now simply $y_{n+N-1}+y_n=0$, i.e. an equation linear in $y$ of order $N-1$. Solving it introduces $N-1$ integration constants and one needs only to integrate the remaining homographic equation for $x$.
\bigskip
5. {\scap Conclusion}
\medskip
Integrability of discrete systems is related to low-growth properties, in particular, for birational systems, to the degree growth of the iterates of some initial condition. This property is encapsulated in the dynamical degree, a value larger than 1 for the latter signalling non-integrability. The practical way to compute the degree of the solution is usually done by obtaining the successive rational expressions and simplifying out the greatest common divisor. This approach, though straightforward, is time-consuming and becomes easily prohibitive, in particular for non-integrable systems. 

Halburd has provided two different approaches aiming at simplifying this task. The first consists in starting from initial conditions that are pure rational numbers, in which case the greatest common divisor simplifications are computationally less heavy. (Even thus, the calculation of the dynamical degree for non-integrable  mappings may turn out to be computationally demanding). The other approach is based on singularities which, as we have explained, are at the origin of the simplifications of the rational expressions. This approach can give the degree of the solution of an integrable mapping. In the case of a non-integrable one, the dynamical degree can be obtained simply using the ``express'' method [\express], 
but the degrees of the solution cannot be obtained in closed form. In this work we showed that by applying Halburd's method of `balancing the singularities' (which is a way of saying that the calculation of the degree growth based on any of the values appearing in the singularity pattern should lead to the same result), the calculation of the degrees can be described algorithmically. Moreover the computations use integer arithmetic with only 
addition, subtraction and, in some cases, multiplication by small integers
and are thus lightning-fast. 
We demonstrated the efficiency of our method by computing the explicit degrees for 
various systems, integrable and not.

In the case of mappings of order higher than two we applied the prescription introduced in [\highorder], whereupon the first initial values of the variables of the mapping are generic and the degree is calculated only in terms of the last one. (For a mapping of order $N$ this means generic values for indices 0 to  $N-2$ and choosing the variable with index $N-1$ as special). 
There is no guarantee that this approach would lead to the correct dynamical degree. However in a slew of cases we studied, some of which were presented here, it turned out that the value of the dynamical degree we obtained coincided with the one given by the Diophantine approach.

Among the examples presented in this paper one mapping was not of birational type. Although the techniques we use have been applied for birational mappings, it turned out that the study of singularities can also be constructively applied to this non-birational case. Whether this can be generalised to other non-birational or even non-algebraic mappings  [\refdef\nobe] is an open question. 

Another intriguing question is that of mappings which, although birational, do not possess rational invariants. The examples of what was called ``solvable chaos''  [\refdef\solchaos] spring to mind. Consider for instance the linear equation $\omega_{n+2}+\omega_{n-2}=2(\omega_{n+1}+\omega_{n-1})$ and introduce a variable $x$ thorough $x=\tan\omega$. Using the addition formulae for the tangent we get the mapping
$$2{1-x_{n+2}x_{n-2}\over x_{n+2}+x_{n-2}}={1-x_{n+1}x_{n-1}\over x_{n+1}+x_{n-1}}-{x_{n+1}+x_{n-1}\over1-x_{n+1}x_{n-1}}.\eqdef\vtes$$
The dynamical degree is $(1+\sqrt 3)/2+\sqrt{\sqrt3/2}\approx2.29663$ and the sequence of degrees, 0,  0, 0, 1, 2, 4, 10, 23, 52, 120, 276, 633, 1454,$\dots$ conforms with this value. The only singularities of the mapping are the two anticonfined ones $\{\cdots,\mp i^{\{2\}},\mp i,\bol,\bol,\bol,\pm i,\pm i^{\{2\}},\pm i^{\{4\}},\pm i^{\{10\}},\cdots\}$ where the upper index (between braces) corresponds to the multiplicity of the root and directly gives the degree growth. 
A domain where the fast algorithm may turn out to be useful is that of lattice systems. Thanks to recent results  [\refdef\mase]  by T. Mase, it is now clear how one can extend Halburd's singularity based method to the case of partial difference equations and obtain the degree growth. We hope to address some of the open problems hinted at in this section in some future work of ours.

\bigskip
{\scap Acknowledgements}

RW would like to thank the Japan Society for the Promotion of Science (JSPS) for financial support through the KAKENHI grants 22H01130 and 23K22401.
\medskip

\bigskip
{\scap References}
\medskip

\begin{description}

\item{[\ince]} E. Hille, {\sl Ordinary Differential Equations in the Complex Domain}, J. Wiley and Sons, New York, (1976).
\item{[\martin]} N.J. Zabusky and M.D. Kruskal, Phys. Rev. Lett. 15 (1965) 240.
\item{[\hirota]} R. Hirota, J. Phys. Soc. Jpn. 45 (1978) 321.
\item{[\cimpa]} B. Grammaticos and A. Ramani, {\sl Integrability of Nonlinear Systems}, Y. Kosmann-Schwarbach, B. Grammaticos, K.M. Tamizhmani (Eds.), Springer LNP 495 (1997) 31.
\item{[\physrep]} A. Ramani, B. Grammaticos, T. Bountis, Physics Reports 180 (1989) 159.
\item{[\sincon]} B. Grammaticos, A. Ramani and V. Papageorgiou, Phys. Rev. Lett. 67 (1991) 1825.
\item{[\dps]} A. Ramani, B. Grammaticos and J. Hietarinta, Phys. Rev. Lett. 67 (1991) 1829.
\item{[\hiv]} J. Hietarinta and C. Viallet, Phys. Rev. Lett. 81 (1998) 325.
\item{[\redem]} A. Ramani, B. Grammaticos, R. Willox, T. Mase and M. Kanki, J. Phys. A 48 (2015) 11FT02.
\item{[\stokes]} A. Stokes, T. Mase, R. Willox and B. Grammaticos, {\sl Full deautonomisation by singularity confinement as an integrability test} (2024) arXiv:2306.01372 [nlin.SI].
\item{[\ablow]} M.J. Ablowitz, R. Halburd and B. Herbst, Nonlinearity 13 (2000) 889.
\item{[\nevan]} B. Grammaticos, T. Tamizhmani, A. Ramani and K.M. Tamizhmani, Intl. J. on Comp. and Math. 45 (2003) 1001.
\item{[\veselov]} A.P. Veselov, Commun. Math. Phys. 145 (1992) 181.
\item{[\arnold]} V. I. Arnold, Bol. Soc. Bras. Mat. 21 (1990) 1.
\item{[\diller]} J. Diller and C. Favre, Amer. J. Math. 123 (2001) 1135.
\item{[\rod]}  R.G. Halburd, Proc. R. Soc. A, 473 (2017) 20160831.
\item{[\dioph]} R.G. Halburd, J. Phys. A 38 (2005) L263.
\item{[\hone]} A.N.W. Hone, Symmetry Integrability Geom. 3 (2007) 22.
\item{[\express]} A. Ramani, B. Grammaticos, R. Willox and T. Mase, J. Phys. A 50 (2017) 185203.
\item{[\highorder]} A. Ramani, B. Grammaticos, A.S. Carstea and R. Willox, {\sl Obtaining the growth of higher order mapping through the study of singularities} (2024) hal-04598791.
\item{[\frompv]} T. Tokihiro, B. Grammaticos and A. Ramani, J. Phys. A 35 (2002) 5943.
\item{[\noncon]} A. Ramani, B. Grammaticos, R. Willox, T. Mase and J. Satsuma, J. Integr. Sys. 3 (2018) 1.
\item{[\thrid]} R. Willox, T. Mase, A. Ramani and B. Grammaticos, Open Communications in Nonlinear Mathematical Physics (2024)  ocnmp:13267.
\item{[\thirdkind]} A. Ramani, Y. Ohta and B. Grammaticos,  Nonlinearity 13 (2000) 1073.
\item{[\gambier]} B. Grammaticos, A. Ramani and S. Lafortune, Physica A 253 (1998) 260.
\item{[\anticon]} T. Mase, R. Willox, B. Grammaticos and A. Ramani, J. Phys. A 51 (2018) 26520.
\item{[\suris]} Y. B. Suris, J. Math. Phys. 37 (1996) 3982.
\item{[\vasfrank]} V. G. Papageorgiou and F. W. Nijhoff, Physica A228 (1996) 172.
\item{[\stef]} A.S. Carstea, J. Math. Phys. 64 (2023) 033504.
\item{[\viallet]} C.M. Viallet, Open Communications in Nonlinear Mathematical Physics, Special Issue in Memory of Decio Levi (2024) ocnmp.11727.
\item{[\lattice]} R. Willox, T. Mase, A. Ramani and B. Grammaticos, J.Phys. A 49 (2016) 28LT01.
\item{[\nobe]} A. Nobe, J. Math. Phys. 65 (2024) 072702.
\item{[\solchaos]} B. Grammaticos, A. Ramani and C.M. Viallet, Phys. Lett. A 336 (2005) 152.
\item{[\mase]} T. Mase, {\sl Exact calculation of degrees for lattice equations: a singularity approach} (2024) arXiv:2402.16206 [nlin.SI].

\end{description}

\end{document}